\documentclass[12pt]{aastex63}

\shorttitle{JAGB Distance Scale IV}
\shortauthors{Madore. Freedman \& Lee}
\graphicspath{{./}{figures/}}

\begin{document}

\title{\bf Astrophysical Distance Scale \\ IV.   Preliminary Zero-Point Calibration of the JAGB Method \\ in the HST/WFC3-IR Broad J-Band (F110W) Filter}


\author[0000-0002-1576-1676]{\bf Barry F. Madore} 
\affil{The Observatories, Carnegie
Institution for Science, 813 Santa Barbara St., 
Pasadena, CA ~~91101, USA}
\affil{Department of Astronomy \& Astrophysics, University of Chicago, 5640 South Ellis Avenue, Chicago, IL 60637, USA}
\email{barry.f.madore@gmail.com} 

\author[0000-0003-3431-9135]{\bf Wendy~L.~Freedman}
\affil{Department of Astronomy \& Astrophysics, University of Chicago, 5640 South Ellis Avenue, Chicago, IL 60637, USA}
\affil{Kavli Institute for Cosmological Physics, University of Chicago,  5640 S. Ellis Ave., Chicago, IL 60637, USA}
\email{wfreedman@uchicago.edu}

\author[0000-0002-5865-0220]{\bf Abigail J. Lee}
\affil{Department of Astronomy \& Astrophysics, University of Chicago, 5640 South Ellis Avenue, Chicago, IL 60637, USA}
\affil{Kavli Institute for Cosmological Physics, University of Chicago,  5640 S. Ellis Ave., Chicago, IL 60637, USA}
\email{abbyl@uchicago.edu}


\begin{abstract} 

We present an absolute calibration of the J-region Asymptotic Giant Branch (JAGB) method using published photometry of resolved stars in 20 nearby galaxies 
observed with HST using the WFC3-IR camera and the  F110W (Broad J-Band)
filter. True distance moduli for each of the galaxies are based
on the Tip of the Red Giant Branch (TRGB) method as 
uniformly determined by Dalcanton et al. (2012). From a composite 
color-magnitude diagram composed of over 6 million stars, leading to a sample of 453 
JAGB stars in these galaxies, we find $M_{F110W}^{JAGB} = -5.77 \pm 0.02$~mag 
(statistical error on the mean). The external scatter seen in a comparison of the individual TRGB and the JAGB moduli is  $\pm$0.081~mag (or 4\% in distance). Some of this scatter can be attributed to small-number statistics arising from the sparse JAGB populations found in the generally low-luminosity galaxies that comprise the particular sample studied here.
However, if this inter-method scatter is shared equitably between the JAGB and TRGB methods that implies that each are good to $\pm$0.06~mag, or better than 3\% in distance.

\end{abstract}



.


\section{Introduction}
Recent measurements of the Hubble constant from direct distance ladder methods (e.g.  Riess et al. 2019) currently differ significantly from values inferred by indirect methods, such as from the modeling of the cosmic microwave background (e.g. Planck Collaboration et al. 2020), although other direct measurements result in better agreement (e.g.,  Freedman et al. 2019, 2020; Freedman 2021). As such, robust tests for systematics are crucial in order to determine whether the current cosmological paradigm needs revision. A new extragalactic distance scale using J-region Asymptotic Giant Branch (JAGB) stars can help break the impasse by providing valuable independent insight into the Hubble tension.

JAGB stars have numerous advantages for measuring distances over the more classical local distance indicator, the Cepheid Leavitt Law, and also over the recently developed Tip of the Red Giant Branch (TRGB) method. First, JAGB stars are distinct and easily identifiable on the basis of their near-infrared colors. They are also ubiquitous, being found in galaxies of all morphological types and inclinations, whereas Cepheids are only found in late-type spiral and irregular galaxies. 
For more distant applications, it is worth noting that the 
JAGB stars are about one magnitude brighter in the near infrared 
than their TRGB counterparts, which have already been successfully
used to determine the local value of the Hubble constant by
calibrating Type Ia supernovae in host galaxies at 
distances of up to 30~Mpc (e.g., Freedman 2021 and 
references therein). The added leverage provided 
by the brighter JAGB stars will make it possible to extend 
the supernova calibration out to significantly larger volumes 
(using HST now, and using JWST in the 
near future) and thereby reaching a significantly larger 
population of SN-host-galaxy calibrators. 

The JAGB method capitalizes on a class of carbon-rich AGB stars, whose evolutionary histories and structure have been well-studied theoretically (e.g., Habing \& Oloffson 2004, Marigo et al. 2008, 2017). In brief, JAGB stars contain an electron-degenerate carbon-oxygen core surrounded by He- and H-fusing shells. These stars are subject to thermal instabilities in their He-shells, leading to `dredge-up' phases, during which the convective envelope of the star penetrates deep onto the He-burning shell, bringing  carbon products up to the stellar surface. 
The resulting carbon on the surface of these now carbon-rich stars gives them a much redder appearance than their bluer oxygen-rich AGB progenitors (Habing \& Oloffson 2004). This process is only effective for stars within a narrow range of masses ($2-5 M_{\odot}$) and ages (300 Myr$-$1 Gyr), and thus a narrow range of absolute luminosities. 
In fact, a review of carbon stars as distance indicators by Batinelli \& Demers (2005) found that the mean luminosity of these stars was constant from galaxy to galaxy, and therefore could be used as a standard candle. More recently, Madore \& Freedman (2020) found that the mean $J$-band luminosity of JAGB stars has zero slope, and Freedman \& Madore (2020) further found that this $J$-band luminosity was relatively insensitive to metallicity. 
Thus, JAGB stars have a well-defined range of luminosities and colors, and therefore are easily identifiable in a near-infrared color magnitude diagram.

This is the fourth in a series of papers exploring the calibration 
and application of the JAGB method to the extragalactic distance scale. 
In Madore \& Freedman (2020; hereafter Paper I) we re-introduced the
method, first used by Weinberg \& Nikolaev (2001, hereafter WN01), and simultaneously
applied by van der Marel \& Cioni (2001), in their differential mapping
of the line-of-sight tilt of the LMC. In Paper I we broadened the
application of the JAGB method more generally to the extragalactic
distance scale, establishing the absolute zero point at the LMC and 
SMC, both of which have independent geometric distances based of
Detached Eclipsing Binary (DEB) stars (Pietrynski et al. 2019 and 
Graczyk et al. 2014, respectively). We then went on to apply the 
JAGB method to a significantly more  distant galaxy, NGC~253, 
obtaining a JAGB distance of 3.40~Mpc, which is close to its TRGB distance 
of 3.46~Mpc (Dalcanton et al. 2009). We note also that a growing number of 
authors (Ripoche et al. 2020; Parada et al. 2021; Zgirski et al. 2021) have  also advocated for the use of this class of carbon stars as extragalactic distance indicators with the latter paper generally confirming the earlier conclusions and calibrations.

In Paper II (Freedman \& Madore 2020)\footnote{See Papers I and II and references therein for a detailed description of the method.} we assembled 
near-infrared data for 16 galaxies having published, 
ground-based, J-band photometry. These data were deep enough to measure, to relatively high precision, the bright, asymptotic giant 
branch (AGB) population of intermediate-aged stars, of which the JAGB 
stars are an IR color-selected subset. In the inter-comparison of the 
JAGB-TRGB distance moduli for this sample, the total observed scatter
(contributed by both methods added in quadrature) was found
to be $\pm$0.08~mag. This  suggested (conservatively)\footnote{Assuming
(unrealistically) that all of the scatter is to be attributed to the 
JAGB distance determinations alone.} that the JAGB method has the potential to measure 4\% distances to individual galaxies.

In Paper III (Lee, Freedman, Madore, et al. 2021) we inter-compared the independently calibrated distances determined by Cepheids, the TRGB method and JAGB stars
in the nearby, Local Group member Wolf-Lundmark-Melotte (WLM), finding agreement, at the 3\% level amongst all three astrophysical methods. We also found that the JAGB method had comparable or lower uncertainties to the Cepheid Leavitt Law and multi-wavelength TRGB, and therefore competitive precision as a distance indicator (see also Zgirski et al. 2021 for an analysis of a complementary sample of galaxies coming to similar conclusions.).

In this Paper IV, we again examine published data, this time exploiting
observations obtained with HST and its WFC3-IR camera, by Dalcanton et al. (2012). This is the first paper to exploit space-based observations for the JAGB method, which we demonstrate is the clear path moving forward if we are to use the JAGB method to measure the Hubble constant and determine distance to galaxies past the current limits of the Cepheid Leavitt Law and TRGB ($\sim 30$ Mpc).

This paper is organized as follows: In Section 2, we describe the HST photometry from Dalcanton et al. (2012) utilized in this study. In Section 3, we recount our methodology for identifying JAGB stars in the 20 galaxies used for the calibration. In Section 4, we present our final F110W calibration and uncertainties. In Section 5, we look to the future for the JAGB method and in Section 6, we summarize this paper's findings and implications. 
\vfill\eject
\section{The HST WFC3-IR Dataset}
For this study, we utilized the publicly available HST WFC/ACS and WFC3/IR photometry for nearby galaxies ($\lesssim 4$ Mpc) from Dalcanton et al. (2009) and Dalcanton et al. (2012), respectively. We note that these data were  taken for other purposes, and hence, not optimized for the measurement of the JAGB.
We acquired the F814W data from Dalcanton et al. (2009), an ACS optical survey of 69 nearby galaxies with derived TRGB distances.
Dalcanton et al. (2012) subsequently supplemented the ACS data with WFC3/IR observations of those galaxies, from which we obtained the F110W data. 
Although WFC3-IR indeed does have a filter (F125W) which is
closely matched to the ground-based J-band filter used in
Papers I and II, many observers (including Dalcanton et al. 2012) have
optioned for a wider NIR filter, F110W, that includes most of the
ground-based J band, but also extends further to the blue, overlapping
(by about 500\AA) with the I-band filter (which is nominally centered around
8000\AA). In maximizing the signal-to-noise ratio of any given observation, choosing the wider 
filter has often won out over the competing desirability of being closer in 
effective wavelength to a more standard, ground-based filter system. 
The F110W filter extends from 9,000 to 14,000\AA, with an effective (pivot) 
wavelength of 11,400\AA. On the other hand, the F125W filter extends 
from 11,000 to 14,000\AA, and has a pivot wavelength of 12,400\AA.
As such, WFC3-IR imaging through the F110W filter can reach a signal-to-noise ratio for K5~III star (say) in an integration time that is 1.6 times faster than exposures using the F125W filter. That advantage is considerable and compelling. We discuss the F160W filter in Section 5.1.

\section{Methodology}

To our knowledge there are no F110W observations of JAGB stars in either the LMC or the SMC, where, in principle, a zero-point calibration could have been obtained for this filter. There are, however, TRGB distances to 20 Dalcanton galaxies being considered here, uniformly determined in Dalcanton et al. (2009). 
We can, therefore, provisionally use the TRGB distances to calibrate, and set the zero point for the JAGB method,
based on the F110W flight-magnitude system. 

The 20 galaxies entering our WFC3-IR (F110W) calibration of the JAGB method are listed in Table 1. The extinction- and reddening-corrected\footnote{ For the J-band (F110W) Milky Way foreground extinctions we are using were taken from the new NED extinction calculator, which are based on Schlafly \& Finkbeiner (2011). 
The average F110W extinction for this sample is 0.038~mag, the median value is 0.03, and the mode is 0.014~mag, with extremes of 0.102~mag, for [HS98] 117, and a low of  0.011~mag for NGC~300.} color-magnitude diagrams (CMDs) are shown individually in Figures 1a and 1b for F814W vs [F475W - F814W] and for F110W vs [F814W - F110W]. Figure 2 shows the distance-corrected composite F110W vs [F814W - F110W] CMD for all 20 galaxies, with the total sample of 6 million stars plotted. The trace of the TRGB, sloping upward at an absolute (F110W) magnitude of about $-4.8$~mag, is shown at a [F814W - F110W] color of about 0.8~mag. Figure 3 shows the same plot but with only one in ten points plotted. This plot shows how clearly and well-defined the TRGB is in this infrared CMD, solidly establishing the zero point for the JAGB stars above it. In the side panels to the right of these two figures we show the marginalized (logarithmic) luminosity functions for all of the  stars in the CMD. These are dominated by the RGB and AGB populations. A linear increase in the AGB luminosity function is shown by a rising white line starting at $M_{F110W} = -7.0$~mag in the right panel, terminating at $M_{F110W} = -5.0$~mag at which point the the TRGB population is first detected. The onset of the TRGB can be traced for about $0.3$~mag, after which the RGB luminosity function dominates and continues rising linearly for about 2 mag below the tip, with a slope of about +0.25 dex/mag. Below $M_{F110W}=-3.0$~mag the apparent luminosity quickly becomes incomplete and rapidly peals away from the projected RGB luminosity function. At brighter magnitudes, the JAGB stars can be seen at $M \sim -5.8$~mag rising above the underlying AGB population.      

The modal J-band magnitude of candidate JAGB stars having $1.50 < [F814W- F110W] < 2.50$~mag in our composite CMD
was found to be $M_{F110W} = -5.77 \pm 0.02 ~(0.37/\sqrt{453}) $~mag (error on the mean). This value, scaled by the (extinction-corrected) TRGB magnitude, was then used to situate the boxes seen in Figure 1a and b. Stars within the boxes seen (having a total F110W magnitude range of $4\sigma =$ 1.4~mag, adopting a single-epoch scatter of $\sigma=0.35$~mag as found in WN01) were identified and their mean and median magnitudes determined. Slight modifications to the box size were made for three galaxies; they are discussed in the notes at the bottom of the table.

 As noted, the Dalcanton galaxies (primarily dwarfs), and the HST flight-magnitude filters chosen to observe them, were not selected by us, nor were they optimized for JAGB detection and/or measurement. 
Most of the galaxies are low-mass systems having only small populations of JAGB stars. Accordingly, in measuring individual distances to galaxies, the modal JAGB magnitude failed to be uniquely identifiable in the very small samples, whereas using the modal magnitude for the composite CMD was found to be most robust as it was the least susceptible to outliers.
Thus, we adopted criteria that matched what we know about the behavior of JAGB stars in the J vs (J-K) CMD and applied those criteria uniformly to the entire sample of 20 galaxies. If there is a figure of merit for these choices it might be had from the quantitative comparison of the mean and median magnitudes of the JAGB candidates on a galaxy-by galaxy basis, and then again from a comparison of these JAGB distances to the TRGB distances to these same galaxies. Comparing the means and median values give $<$mean – median$>$ = +0.009~mag, which is consistent with the underlying parent populations being highly symmetric (but not necessarily Gaussian), as we know to be the case from the LMC and SMC samples. In the second comparison, as Figure 4 shows, the agreement between the JAGB distance scale (as defined by the adopted procedures) is in excellent systematic agreement with the TRGB distance scale: that being  $<$TRGB – JAGB (mean)$>$  =  +0.008 $\pm$ 0.080~mag [sigma on the mean = $\pm$0.018~mag] and $<$TRGB - JAGB (median)$>$ = -0.002 $\pm$0.090 [sigma on the mean = $\pm$ 0.021~mag]. The mean and median values for the absolute magnitudes of the JAGB populations studied here are found to be -5.74 and -5.76 mag, respectively. 
These agreements were by no means guaranteed by anything adopted in our procedures, but they suggest that they are not unreasonable choices. Attempting to go further in adjusting the parameters, we believe, would be pushing these datasets too hard. For the findings made by others grappling with these same problems, we refer the reader to the papers by Parada et al. (2021) and Zgirski et al. (2021).

We thus conclude that even under sub-optimal conditions (of sample size, non-standard filters and non-standard color combinations, and in the presence of asymmetric contamination due to unresolved background galaxies) the JAGB method appears to be robust.

\begin{figure}[hbt!]
\figurenum{1a}
\centering
\includegraphics[width=15.9cm, angle=-0] {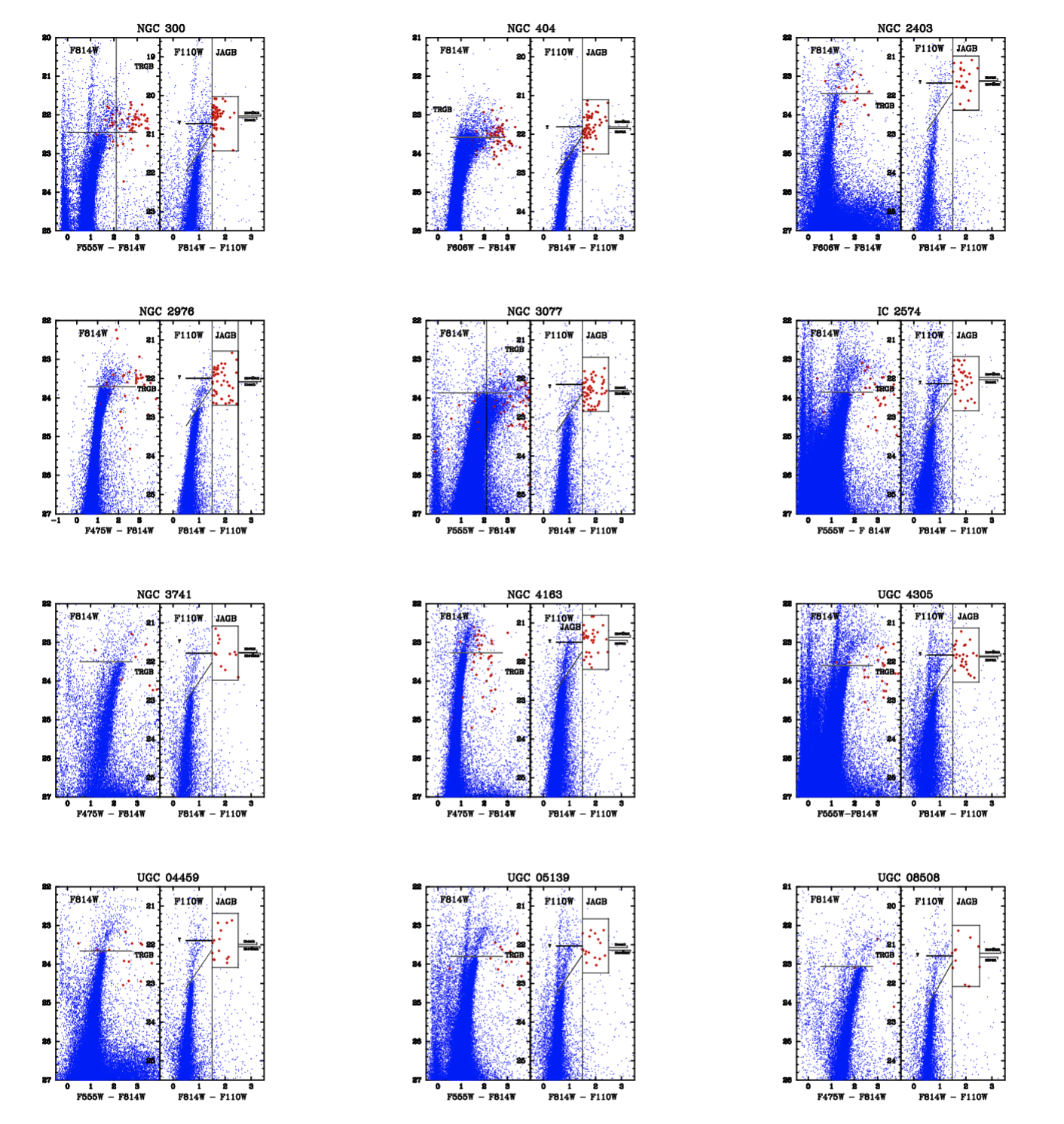}

\caption{\small -- Individual color-magnitude diagrams for each of the 20 galaxies discussed in this paper.  The left sub-panel is the F814W vs (F606W - F814W) or (F475W - F814W) or (F555W - F814W), CMD where the constant level of the I-band (F814W) magnitude of the TRGB (as given in Dalcanton et al. 2009) is marked by a horizontal solid black line. The right sub-panel is the F110W vs [F814W - F110W] CMD in which the JAGB population is identified, shown as circled dots found within the rectangle to the right of the vertical line at $[F814W - F110W] =$ 1.5 mag.}
\end{figure}

\begin{figure}[hbt!]
\figurenum{1b}
\centering
\includegraphics[width=15.9cm, angle=-0] {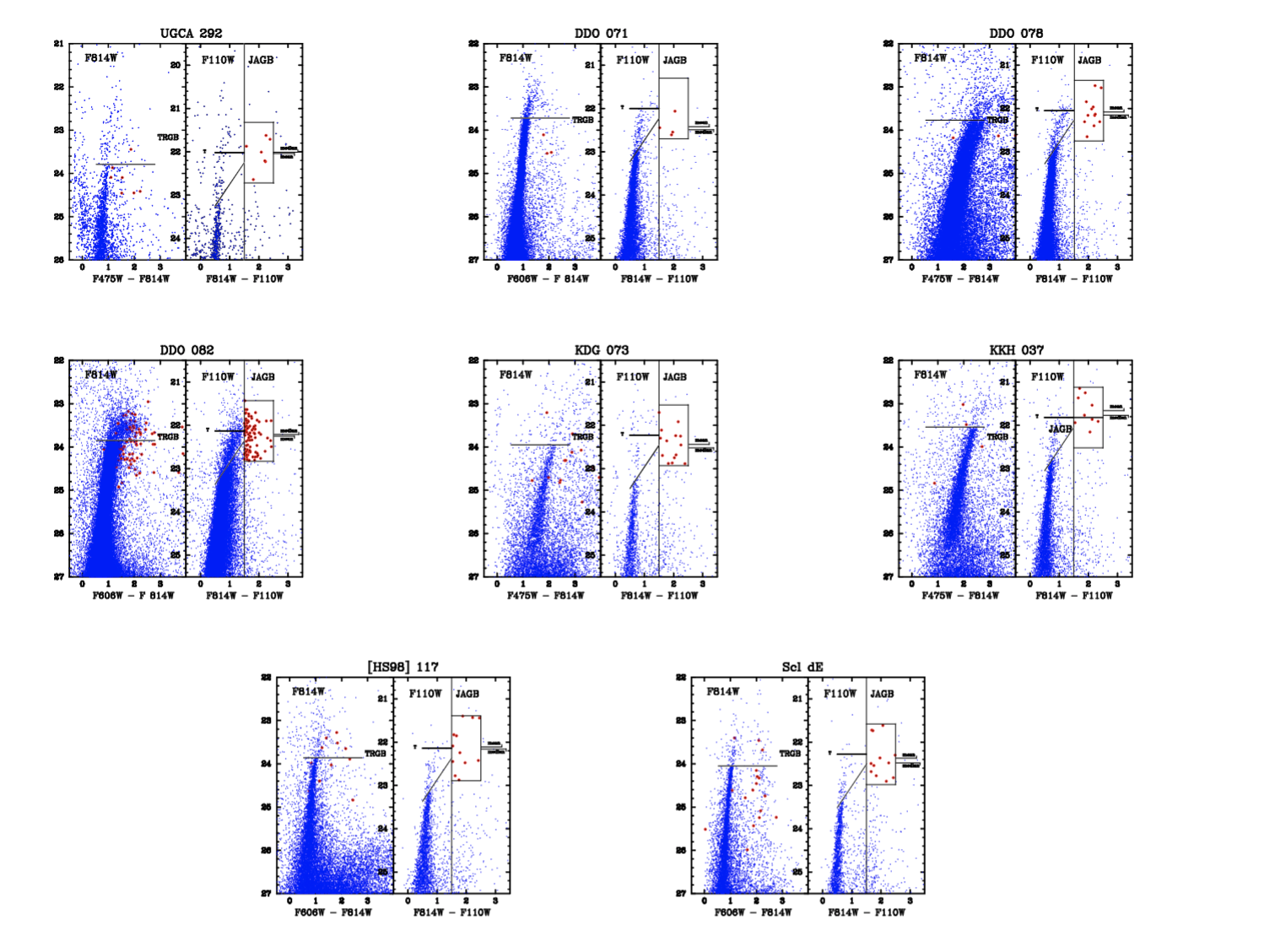}

\caption{\small --Same as Figure 1a.}
\figurenum{1b}
\end{figure}

\begin{figure}[hbt!]
\figurenum{2}
\centering
\includegraphics[width=15.9cm, angle=-0] 
{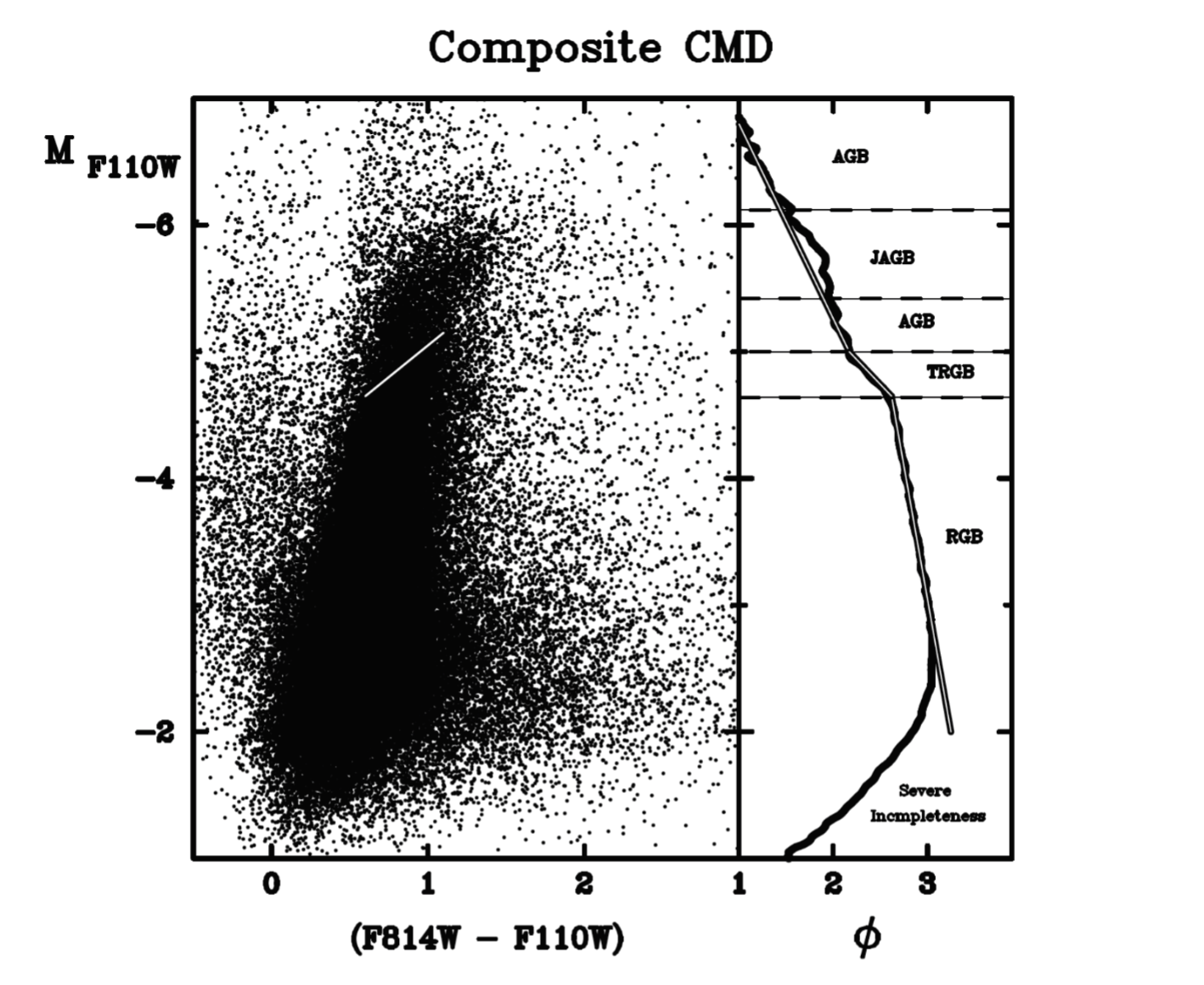}
\caption{\small -- The composite color-magnitude diagram based on over 6 million stars in the 20 galaxies discussed in this paper. The left-hand panel shows all of the points, emphasizing the AGB population (above the slanting white line, which marks the position of the TRGB, more easily seen in Figure 3). The marginalized (logarithmic) luminosity function for this composite CMD is seen in the panel to the right. From top to bottom, the white lines show the power-law rise of the AGB population from $M_{F110W} \sim -7$~mag to $-5$~mag, at which point the upward-slanting TRGB can be seen, transitioning to a second power-law rise, defined by the Red Giant Branch (RGB) population. At intermediate luminosities, around $M_{F110W} \sim -6$ mag the JAGB population can be seen superimposed upon  the general population of O-rich AGB stars. }  
\end{figure}

\begin{figure}[hbt!]
\figurenum{3}
\centering
\includegraphics[width=15.9cm, angle=-0]{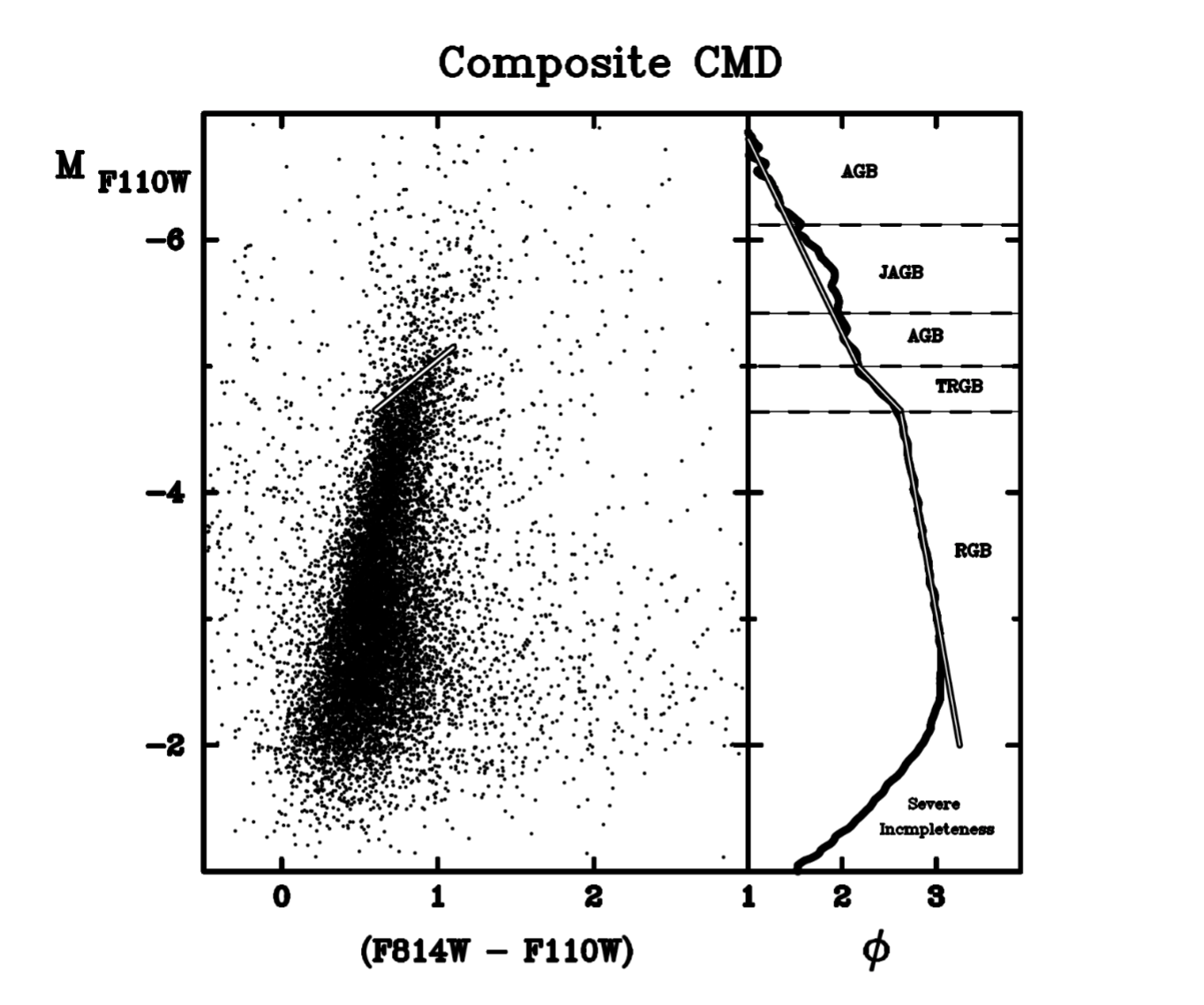}
\caption{\small -- Same as Figure 2, except that only one in ten stars have been plotted so as to allow the trace of the TRGB (upward-slanting white line at -5 mag) to be readily seen.}
\end{figure}

\vfill\eject

\medskip\medskip\medskip\medskip\medskip\medskip\medskip\medskip
~
\section{Results}

Table 1  summarizes the TRGB and JAGB distances for the 20 galaxies in this study, based on the CMDs presented in Figures 1a and 1b. The first column gives the name of the galaxy followed by the TRGB true distance modulus and its error, taken without modification from Dalcanton et al. (2012). Columns 4 and 5 give the JAGB true moduli and their statistical uncertainties. The number of stars, $N_J$, entering the distance estimation are given in Column 6. The errors on the JAGB distance moduli are calculated from prior knowledge of the JAGB intrinsic dispersion for a single-epoch observation of this population of stars. We adopt a dispersion of $\pm0.35$~mag from Weinberg \& Nikolaev (2001) and calculate the error on the mean by dividing by $\sqrt{N_J-1}$. Most of our JAGB luminosity functions are Poisson-noise-limited in their total counts of JAGB stars (given in Column 6); using their observed scatter would, in many cases, significantly {\it underestimate} the calculated uncertainty.

The individual JAGB distance moduli are compared with their corresponding
TRGB distance moduli in Figure 4.  The solid lines flanking the one-to-one correspondence (dashed) line are at $\pm 2 \sigma$,
where the dispersion was measured internally from the 18 galaxies to be $\sigma = \pm0.080$~mag (having omitted KDG~073 and DDO~071, the two galaxies that deviate most from the 1:1 line). This can be compared to the scatter of $\pm$ 0.08~mag found in the ground-based data (Freedman \& Madore 2020).  In Figure 5 we show the WFC3-IR sample added to the ground-based sample covering a 10~mag span of distance moduli. The diagonal line is not a fit to the data, but a line of slope 1, showing the excellent agreement between the TRGB and JAGB methods. The residual plot, given in the bottom panel of Figure 4, shows no indication of a slope or of any obvious zero-point offset between the near and far samples. The only trend is the slightly increasing scatter with distance, as noted above.

\begin{table}
\centering
\scriptsize
\caption{\bf TRGB and JAGB [F110W] Distances}
\label{tab1}
\begin{tabular}{lcccccccrl}
\hline 
Name &  TRGB& & ~~ & JAGB & & JAGB\cr 
 & $\mu_o$ &$\sigma_m$ & ~~ &$\mu_o$(mean)  & $\sigma_m$ &$\mu_o$(median)&$\sigma_m$&No.&$W_{1/2}$\cr 
~~  & (mag) & (mag) & ~~ &(mag) & (mag)& (mag)& (mag)&Stars&(mag)\cr
\hline 
NGC 0300    & 26.42  & 0.030 && 26.34 & 0.046 &26.30& 0.023& 62 &0.70\cr
NGC 0404~(a)& 27.58 & 0.021 & & 27.58 & 0.044 & 27.63&0.021&68&0.70 \cr 
NGC 2403    & 27.45  & 0.036 && 27.38 & 0.090 & 27.41&0.064 &22 &0.70\cr 
NGC 2976    & 27.76  & 0.034 && 27.87 & 0.051 & 27.86&0.030&46 &0.70\cr 
NGC 3077    & 27.92  & 0.009 && 28.07 & 0.045 &28.07 &0.026&61 &0.70\cr \\
NGC 3741   & 27.55  & 0.023 && 27.53 & 0.111 &27.55& 0.117&13 &0.70\cr 
NGC 4163   & 27.28  & 0.014 && 27.22 & 0.064 &27.14& 0.045&31 &0.70\cr 
IC 2574     & 27.89  & 0.019 && 27.80 & 0.064 &27.74&0.045& 31 &0.70\cr 
UGC 04305   & 27.65  & 0.071 && 27.65 & 0.067 &27.62&0.050& 28 &0.70\cr 
UGC 04459   & 27.79  & 0.022 && 27.76 & 0.094 & 27.82&0.093& 14 &0.70 (b)\cr \\
UGC 05139   & 27.95  & 0.029 && 27.84 & 0.085 &27.91&0.082& 17 &0.70\cr 
UGC 08508   & 27.06  & 0.024 && 27.09 & 0.143 &26.99&0.127&  11 &0.79\cr 
UGCA 292    & 27.79  & 0.022 && 27.81 & 0.143 & 27.78&0.200&  7 &0.70\cr 
DDO 071     & 27.74  & 0.064 && 28.18 & 0.202 &28.26&0.350&  4 &0.70\cr 
DDO 078     & 27.82  & 0.027 && 27.85 & 0.094 &27.92&0.117& 12 &0.70\cr \\
DDO 082     & 27.90  & 0.027 && 28.02 & 0.038 &27.98&0.016& 88 &0.70\cr 
KDG 073     & 28.03  & 0.045 && 28.21 & 0.097 &28.29 &0.100& 14 &0.70\cr 
KKH 037     & 27.57  & 0.026 && 27.43 & 0.124 &27.54&0.156&  9 &0.70\cr 
[HS98] 117  & 27.91  & 0.025 && 27.88 & 0.132 &27.93&0.117& 12 &0.75 (c)\cr 
Scl dE      & 28.11  & 0.037 && 28.14 & 0.106 &28.25&0.117& 12 &0.70\cr 
 \\
\hline \\
\end{tabular}
\tablecomments{ (a) Saturated central regions omitted. (b) UGC~04459: If the window had been increased by 0.02~mag	in order to accommodate the one star at the top
of the interval	the mean and median distance moduli would have become 27.76 and  27.82~mag, respectively, illustrating the sensitivity of the median to small-number statistics, as is the case here.  
(c) [HS98] 117: By increasing the window by $\pm$0.05 mag the number of JAGB stars included in the calculation increases by 50\%.
With the default window	of $\pm$0.70~mag the mean and median distance moduli for the 8 stars
contained in that selection are 28.10 and 27.80~mag, respectively.}
\end{table}


\begin{figure}[hbt!]
\figurenum{4}
\centering
\includegraphics[width=16.0cm, angle=-0] 
{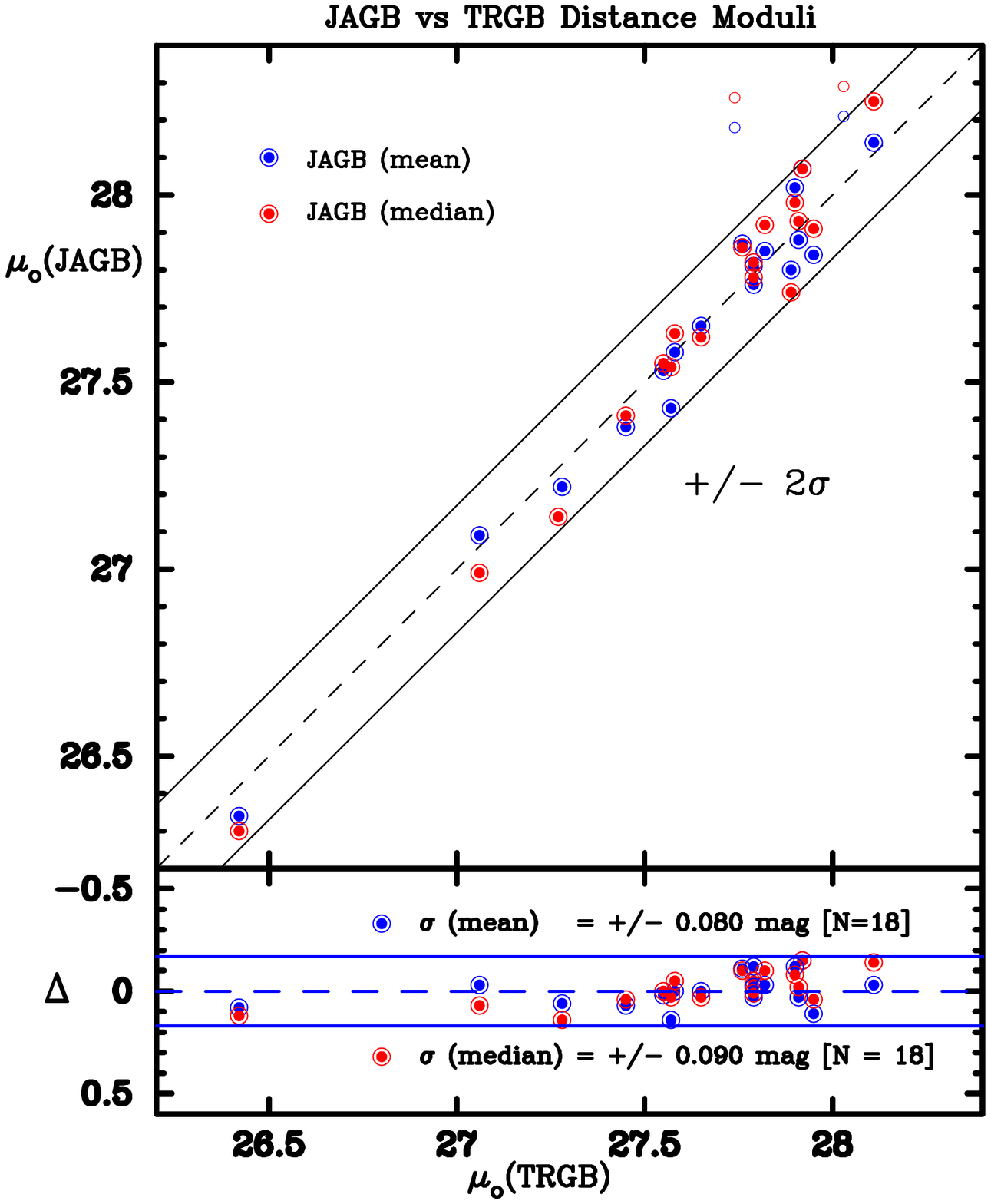}
\caption{A zoomed-in view of the comparison of the TRGB and JAGB distance for the subset of 20 galaxies presented in this paper. See Figure 5 for the total sample of data available to date. The dash-dot lines, flanking the one-to-one (solid blue) line, represent the 2-$\sigma$ scatter as calculated from the plotted points. 
Two galaxies, KDG 073 and DDO 071 fall outside of the two-sigma boundaries of the one-to-one correlation with the TRGB distances. They are flagged as open circles in the plot, and they have been dropped from further consideration in the text}
\end{figure}

\begin{figure}[hbt!]
\figurenum{5}
\centering
\includegraphics[width=20.9cm,angle=-0]{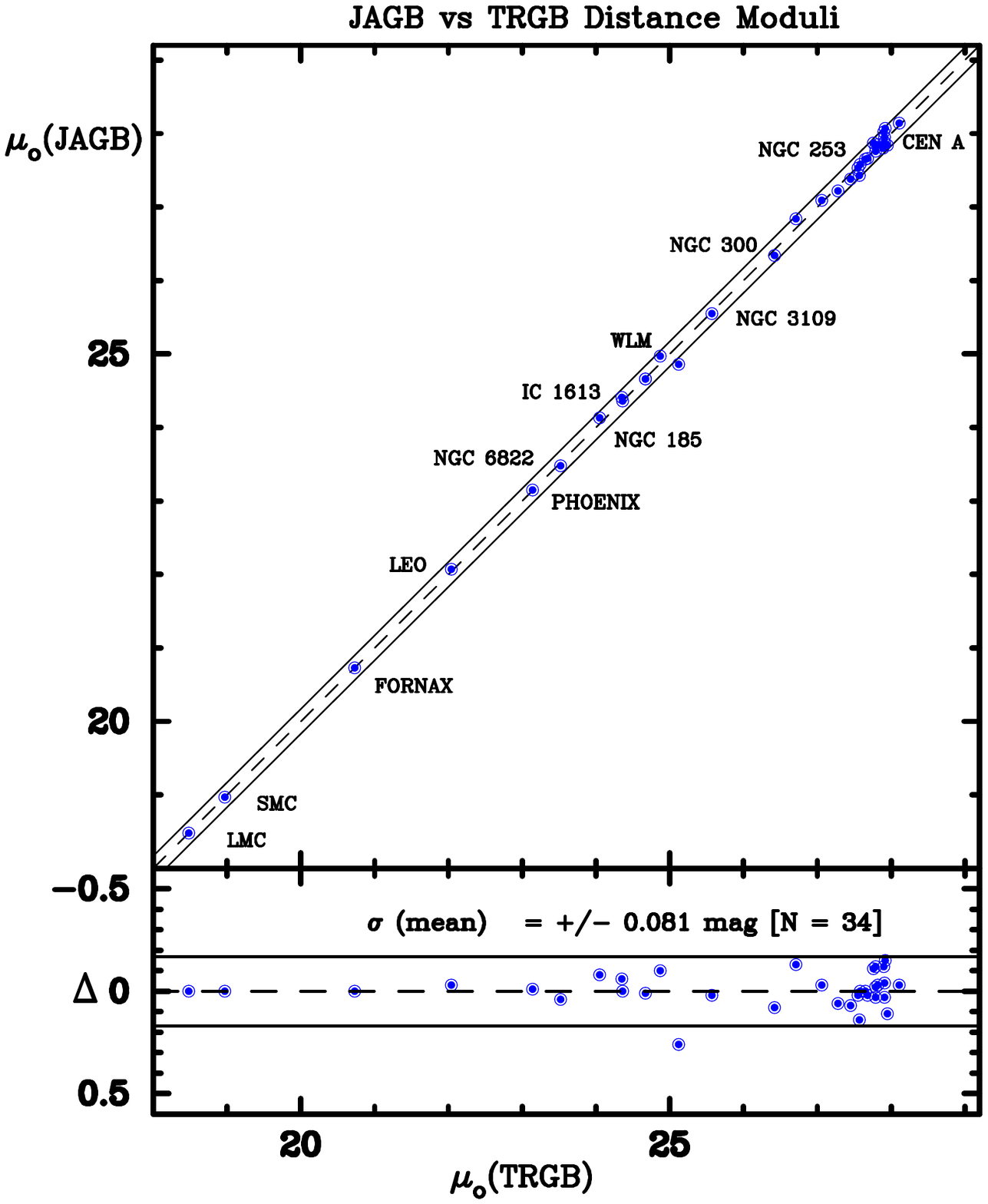}
\caption{\small --  Comparison of the JAGB and TRGB true distance moduli for 34 nearby galaxies (having omitted DDO 71 and KDG 073). The solid black line is a unit slope line passing through (0,0); it is not a fit to the data. Here we have combined two samples  of JAGB distance moduli: a  more distant sample (from  the current paper using HST data), and a more nearby sample (from Freedman \& Madore 2020, based on ground-based data). The total scatter about the unit-slope line is $\pm$0.081~mag. The individual deviations are shown as a function of the TRGB distance modulus in the lower panel, where the individual JAGB error bars are visible. The two horizontal dashed lines mark the two-sigma bounds of the plotted residuals.}
\end{figure}\begin{figure}[hbt!]
\end{figure}

\vfill\eject
\section{Looking to the Future}

Even without having optimized the filter choice, the exposure times and/or numbers of epochs, the results so far (from the ground in Paper II, and now from space in this study) auger well for using the JAGB method to obtain accurate and high-precision distances to galaxies of almost all morphological types. Here we summarize what is known about the JAGB population, with an eye on how best to go forward with future observations.

\medskip

\noindent (1) It is an advantage to apply the  JAGB method in the near infrared at a
wavelength as close as possible to the ground-based, 1.2 micron (J band) 
where the mean absolute magnitude of the JAGB population is independent of color. At shorter or longer wavelengths the run of magnitude 
with color will manifest itself in a broadening of the marginalized luminosity function, or alternatively,
the calibration will require higher-precision colors to de-trend 
the data, thereby requiring additional observing time.

\medskip
\noindent
(2) In the J band the observed (single-epoch) dispersion in
the luminosity function is about $\pm$0.35~mag (Weinberg \&
Nikolaev 2001). 

\medskip
\noindent
(3) The JAGB stars are all expected to be intrinsic variables,
both periodic and irregular, with J-band amplitudes of about
0.9~mag (e.g., Huang et al. 2019). Items (2) and (3) are coupled. The dispersion seen in a
single-epoch observation is the quadrature sum of scatter in the
time-averaged magnitudes of the JAGB stars and their intrinsic
variability, effectively acting as a stochastic variable of bounded amplitude. Since 
the light curves of many AGB stars are closely sinusoidal in
the near infrared, a peak-to-peak amplitude of 0.9 mag would
then effectively translate into a ``random'' error of $\pm$0.32~mag; but this is very close to the total scatter of 0.35 mag seen by Weinberg \& Nikolaev (2001). Indeed, if $\Delta$J = 1.0 mag, then $\sigma$ = 0.35 mag, which would mean that after averaging over the light curves there would be no intrinsic width left over for the JAGB stars seen at their mean magnitudes, an outcome that would be unprecedented, and also very unlikely. 


For our back-of-the-envelope calculation above, we  assumed a marginalized magnitude distribution that is rectangular, in which case for $\Delta$J = 1.0~mag the dispersion is equal to 1.0/$\sqrt{(12)}$ = 0.29 mag. So, reverting to  $\Delta$J = 0.9 mag, we then find $\sigma$ = 0.26 mag, leaving 0.23 mag for the intrinsic scatter, which is confirmed by the scatter remaining in the Macri et al. (2015) data, which were time-averaged over the window used to search for Cepheids in these fields (see also Freedman \& Madore 2020). The other way to proceed, if a sine-wave model is preferred, is to adopt $\Delta$J = 0.75 mag giving $\sigma$ = 0.26 mag, and again leaving 0.23 mag for the intrinsic (star-to-star) scatter.

The above calculations are being shown primarily to point out that both increasing the sample size and/or independently undertaking time averaging can help bring the final variance down at approximately the same rate.
For instance, half of the dispersion can
be averaged out (declining as $\pm~0.26/\sqrt{N_{epochs}}$)
simply by taking multiple observations of the same population, 
randomly spread out over an interval of 100 days or more. Alternatively, the impact of the intrinsic
dispersion on the error on the mean can also be reduced by 
independently observing larger samples of stars in a given galaxy
(until the host galaxy population is fully sampled).
This error on the mean declines as $\pm~0.23/\sqrt{N_{JAGB}}$.

\medskip
\noindent
(4) Ideally the photometric precision of the individual observations 
of the JAGB stars should be kept at, or below, the $\pm$0.25~mag
level so that it does not become a major contributor to the final
error budget on the precision. In the limiting application of this
method to the most distant galaxies, if the on-target integration
needed to obtain a SNR of 10:1 for a given JAGB star (i.e.
$\sigma = \pm$~0.1~mag) requires many orbits then it would be
useful to consider splitting each of those orbits into
separate visits, so as to average over intrinsic variability,
while ultimately collecting the requisite total number of photons.
While this strategy might work well for HST, it is less
obvious for JWST where the target-acquisition overheads are considerably larger; in that case dwelling longer in a single visit may well be the best strategy.

\subsection{Other Considerations}
\par
\noindent
(1) Since the JAGB population is color-selected in the J
band, a second (red) filter is necessary, {\it with one major exception}:
the F160W band on HST/WFC3 is both very narrow and shifted to
the blue of its ground-based H-band equivalent resulting in it falling
on strong molecular features in the spectra of carbon stars. As a result
the color separation of C-rich AGB stars from O-rich AGB stars 
fails completely (see Dalcanton et al. 2012 for a detailed description 
of this unfortunate circumstance). 
However, for other filter choices for a population-discriminating color, the larger the wavelength
separation, the better the color separation will be, at the
same SNR in the second band.
\par
\noindent
(2) The highest surface brightness, inner regions of
galaxies need to be avoided. There are three reasons for
doing so:
\par\noindent
(a) Crowding by other stars simply scales with the
surface brightness. 
\par\noindent
(b) Dust and gas in the inner disk will
systematically dim the magnitudes of the JAGB stars;
this complication can be ameliorated by moving further to the
infrared, but it cannot be completely eliminated, and it is best to avoid
those regions from the outset. 
\par\noindent
(c) The ratio of carbon-rich, C-type AGB stars to normal, O-type AGB stars increases with
decreasing metallicity (e.g., see Brewer et al. 1995). Observing in the 
outer extended disks of spiral galaxies avoids crowding, 
minimizes dust, and because metallicity generally declines 
with galactocentric radius, this acts to increase the relative 
number of JAGB stars with respect to the overall AGB population in these outer regions.

\vfill
\eject
\section{Conclusions}
\medskip
\par
\noindent

Based on a published set of data from HST/WFC3, we find that the JAGB method for measuring high-precision distances to nearby galaxies continues to show considerable promise.  We emphasize that these data were not taken with the purpose of measuring the JAGB, and that further optimization (choice of fields, filters, increase in sample size, etc.) can improve these measurements.

\medskip
\par
\noindent

Using the TRGB distances, independently determined for each of the 20 galaxies published previously by Dalcanton et al. (2012), we have measured the average J-band absolute magnitude of the JAGB stars in the F110W flight-magnitude system, and determined distances based on the JAGB method.

\medskip
\par
\noindent

 In a comparison of the previously-published TRGB distances with the
distances determined in this paper, using the JAGB method, we 
find an inter-method scatter of   $\pm$0.081~mag (or $\pm$~4\% in
distance per galaxy). This is competitive in precision to that
found for supernova-calibrating Cepheid PL distances ($\pm$0.08~mag,
Riess et al. 2016, their Table 5) and for nearby Type Ia supernovae
individually (e.g., $\pm$0.10-0.18~mag, Burns et al. 2018 and
$\pm$0.13~mag, Riess et al. 2016 Table 5). It is worth pointing out that all three of these quoted uncertainties are larger than the commonly quoted precision
of the TRGB method (e.g., $\pm$0.06~mag Rizzi et al. 2005, their Table 5).  We note that if the inter-method scatter of $\pm$0.081~mag, found here, is shared equally between the two methods, they would each have a precision of $\pm0.06$~mag (or 3\% in distance).  

\medskip
\par
Using HST and WFC3-IR it is possible to measure the JAGB population
at a signal-to-noise ratio of 5 (i.e., $\pm$0.20~mag) in a
single-orbit exposure of 2,400~sec for a galaxy at a distance 
of 38~Mpc (or, to the same SNR, out to 60~Mpc in under 6.5 orbits). 
\medskip\par
For JWST we estimate that in 5 hours, distances to
galaxies can be measured with the JAGB technique out to 100~Mpc, 
providing both a valuable calibration of Type~Ia SNe, and an independent 
measure of the Hubble constant at that redshift. For the same integration time, 
but aiming for a signal-to-noise of 10, JWST could obtain JAGB
distances for any galaxy within 70~Mpc, which is five times larger in volume than what is currently within reach of HST using Cepheids. 

\section{Acknowledgements}  
We sincerely thank the referee for very constructive questions and suggestions. We especially appreciate the suggestion to consider more closely the issue of contamination of the JAGB luminosity function by the ever-present (and sometimes unresolved) background galaxy population.

We thank the {\it University of Chicago} 
and {\it Observatories of the Carnegie Institution for Science} 
and the {\it University of Chicago} for their support of 
our long-term research into the calibration and determination 
of the expansion rate of the Universe. Support for this work was 
also provided in part by NASA through grant number HST-GO-13691.003-A 
from the Space Telescope Science Institute, which is operated 
by AURA, Inc., under NASA contract NAS~5-26555. This research was 
greatly enabled by the NASA/IPAC Extragalactic Database (NED), 
which is operated by the Jet Propulsion Laboratory,
California Institute of Technology, under contract with the 
National Aeronautics and Space Administration.
Finally, we thank Julianne Dalcanton for making her HST photometry publicly available.

\section{References}
\medskip
\noindent
Battinelli, P., \& Demers, S.\ 2005, \aap, 434, 657

\noindent{Bershady, M.A., Lowenthal, J.D. \& Koo, D.C. 1998, \apj, 505, 50 }

\noindent
Brewer, J.P., Richer, H.B., \& Crabtree, D.R. 1995, AJ, 109, 2480

\noindent
Burns, C.R., Parent, E., Phillips, M.M., et al. 2018, ApJ, 869, 56

\noindent
Dalcanton, J.J., Williams, B.F., Seth, A.C., et al. 2009, ApJS, 183, 67

\noindent
Dalcanton, J.J., Williams, B.F., Melbourne, J.L., et al. 2012, ApJS, 198, 6 

\noindent
Freedman, W.L, 2021, ApJ, 919, 16 arXiv: 2106.15656

\noindent
Freedman, W.L., Madore, B.F., Hatt, D., et al. 2019, ApJ, 882, 34

\noindent
Freedman, W.L., \& Madore, B.F. 2020, ApJ, 899, 67 

\noindent
Graczyk, D., Pietrzynski, G., Thompson. I., et al. 2014, ApJ, 780, 59

\noindent
Habing, H.J., \& Olofsson, H. 2004, ``Asymptotic Giant Branch Stars'',  Springer-Verlag, New York.

\noindent
Huang, C., Riess, A.G., Hoffmann, S.L., et al, 2018, ApJ, 857, 67

\noindent
Lee, A. J., Freedman, W. L., Madore, B. F., et al. 2021, ApJ, 907, 112.

\noindent
Macri, L., Ngeow, C.-C., Kanbur, S., Mahzooni, S., \& Smitka, M.T. 2015, AJ, 149, 117

\noindent
Madore, B.F., \& Freedman, W.L. 2020, ApJ, 899, 66 

\noindent
Marigo, P., Girardi, L., Bressan, A., et al. 2008, A\&A, 482, 883.

\noindent
Marigo, P., Girardi, L., Bressan, A., et al. 2017, ApJ, 835, 77.

\noindent
Parada, J., Heyl, J., Richer, H., et al. 2021, \mnras, 501, 933

\noindent
Pietrzynski, G., Graczyk, D., Gallenne, A., et al., 2019, Nature, 567, 200

\noindent
Planck Collaboration, Aghanim, N., Akrami, Y., et al., 2020, \aap, 641, A6. 

\noindent
Riess, A. G., Casertano, S., Yuan, W., et al. 2019, ApJ, 876, 85

\noindent
Riess, A.G., Macri, L.M., Hoffmann, S.L., et al. 2016, ApJ, 826, 56

\noindent
Ripoche, P., Heyl, J., Parada, J., \& Richer, H. 2020, MNRAS, 495, 2858
 
\noindent
Schlafly, E.F., \& Finkbeiner, D.P. 2011, ApJ, 737, 103  
 
\noindent
Tikhonov, N.A., Galazutdinova, O.A., \& Aparicio, A. 2003, A\&A, 410, 863
 
\noindent
van der Marel, R., \& Cioni, M.-R., 2001, AJ, 122, 1807

\noindent
Weinberg, M. D., \& Nikolaev, S. 2001, ApJ, 548, 712

\noindent
Zgirski, B., Pietrynski, G., Gieren, W., et al. 2021, ApJ, 916, 19 

\vfill\eject

\appendix

\section{Background Galaxy Contamination}

As emphasized by the referee, one source of contamination of the JAGB population consists of background galaxies falling in the same magnitude and color range as the bona fide JAGB stars.

We offer here several paths forward in evaluating and minimizing the impact of these sources on the derived JAGB luminosity functions and the distance moduli derived from them:

(1)	The background contaminants will come in two types, (a) unresolved point sources and (b) those that are large enough in angular size (for a given telescope and detector) to show structure in the form of nuclei, extended disks and/or halos, ellipticities and other forms of measurable deviations from being point sources. These latter objects can then be automatically flagged (using sharpness and ellipticity criteria, etc.) or found by visual inspection, and then removed from the counts. This would reduce, but not totally account for the background-galaxy contamination. Clearly this procedure will be more effective on space-based images where the resolution is generally far superior to ground-based images. 

(2)	Virtually all of the sources in the color range spanned by the JAGB stars, and fainter than the true JAGB population, will be background galaxies. One can then use the observed luminosity of these fainter sources in a given field to normalize an extrapolation back into the JAGB region so as to estimate the contamination appropriate to that pointing.  The right panel of Figure A1 shows one such slope (N = 0.35) derived {\bf by us} from the Bershady et al. (1998) data, which appears to be a reasonable approximation to the subset of galaxies in the JAGB zone.

\begin{figure}[hbt!]
\figurenum{A1}
\centering
\includegraphics[width=13.9cm, angle=-00] 
{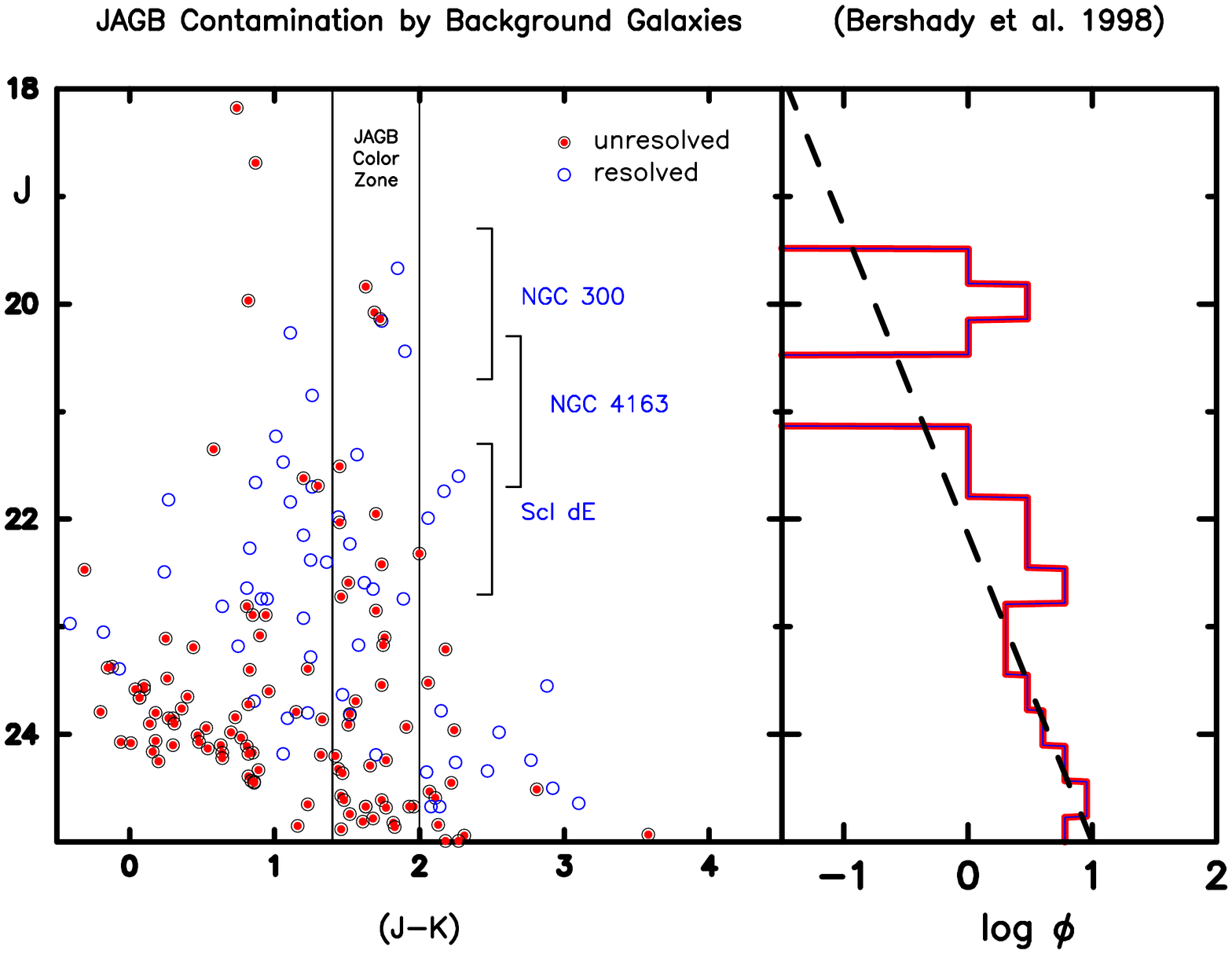}
\caption{\small -- Left Panel -- A composite J vs (J-K) color-magnitude diagram for resolved (red circled dots) and unresolved  (blue open circles) of moderate-redshift galaxies in two blank fields. The two vertical solid black lines at (J-K) = 1.4 and 2.0 mag mark the color limits of the J-Branch AGB (JAGB) stars as defined by Weinberg \& Nikolaev (2001). The three right-angle brackets to the right of the data make the apparent magnitude ranges occupied by JAGB stars in galaxies at distance moduli of 26, 27 and 28 mag, exemplified by NGC~300, NGC 4163 and the Sculptor Dwarf Spheroidal, respectively.  All data are from  Bershady et al. (1998).} The panel to the {\bf right} shows the binned logarithmic luminosity function of the stars in the left panel inside the JAGB color zone. The fit is noisy, but satisfactory. The dashed black line has the slope of the entire galaxy population as given by Bershady et al. See text for additional details. 
\end{figure}

(3)	If available in the discovery frames (or in a local control field) one could measure the contamination directly at the same magnitude and color range as the JAGB population, and apply statistical corrections to the binned JAGB data, without renormalization or extrapolation.

(4)	Finally, published deep-field luminosity functions of faint galaxies over the sky can be used to predict the contamination.
Figure A2 shows one such plot of a J vs (J-K) CMD of galaxies in the magnitude and color range covering the Dalcanton JAGB galaxy fields, using tabulated data in Bershady et al. (1998). That study covered an area of 1.6 square arcmin on the sky which is about one third of the area of WFC3-IR array. Using these data we conclude that background galaxies contribute only 9 to 17 contaminants for the bulk of the galaxies in this study, i.e., those having distance moduli in range of 27.5 to 28.0 mag.

\begin{figure}[hbt!]
\figurenum{A2}
\centering
\includegraphics[width=15.9cm, angle=-00] 
{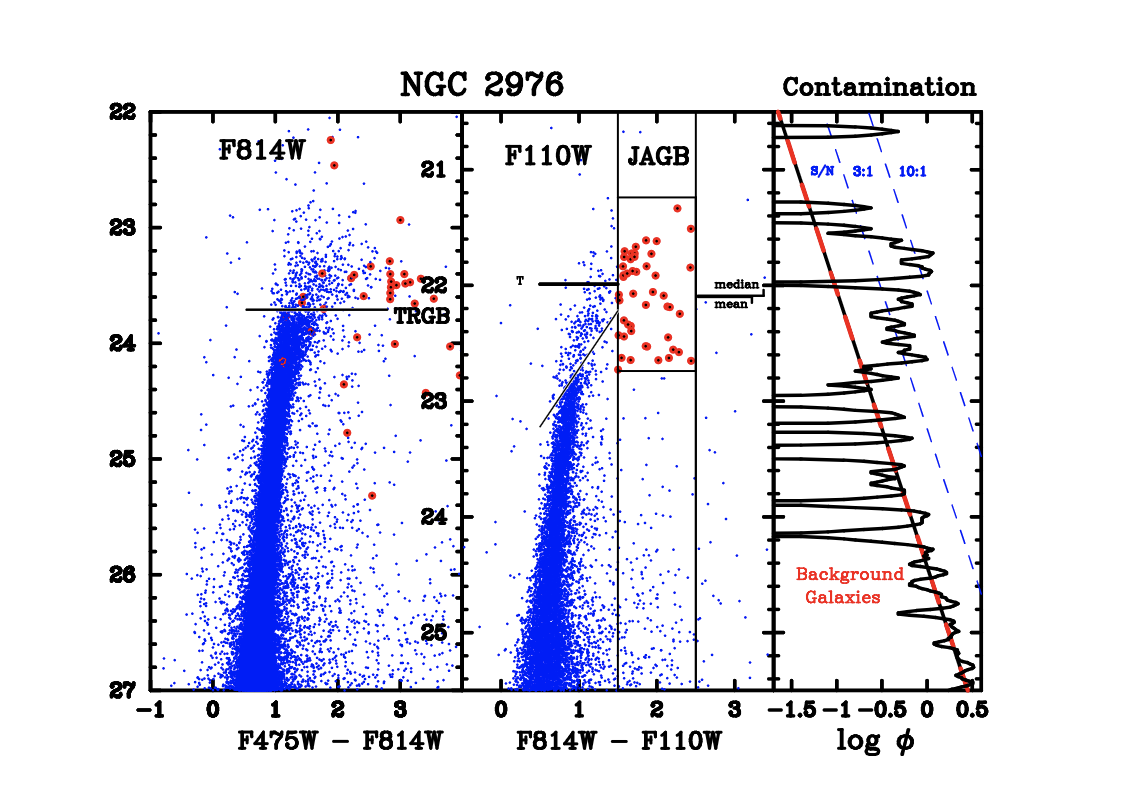}
\caption{\small -- An expanded view of the color-magnitude diagrams for NGC 2976 with the addition of  the right panel that contains the complete J-band logarithmic luminosity function over the entire magnitude range for the color-restricted sampling corresponding to the selection function for JAGB stars [1.5 $<$ (F814W-F110W) $<$ 2.5 mag]. Background galaxies fall below the red-black dashed line, and JAGB stars fall above. Dashed blue lines indicate the ratio of JAGB stars to background galaxies expressed as a signal-to-noise ratio, S/N.
}
\end{figure}
\vfill\eject
We close by noting that the calibrating JAGB J-band luminosity function, as measured in the LMC, is centrally peaked, symmetric and closely Gaussian in form. The fact that many of the more distant galaxies show an extended faint tail to the JAGB luminosity function, leads us to conclude that this feature is plausibly due to unresolved background galaxy counts, falling in the same range of color as used to select JAGB stars at brighter magnitudes. 

Figure A2 illustrates the above contamination effect for one of our galaxies, NGC~2976, at a representative distance modulus of about 27.9 mag.
In the right-most panel, the logarithmic luminosity function is shown for all of the stars in the middle CMD falling in the color range 1.5 to 2.5~mag, as used to select JAGB stars in this paper. A fit to the lower 2.5 magnitudes of the fainter objects is shown by the red-black dashed line, which has been
extrapolated to brighter magnitudes underneath the JAGB population. The background-galaxy luminosity function has a slope of N = 0.42, consistent with the independently determined values of 0.35 derived from the Bershady et al. (1998) data, discussed above and shown in Figure A1. In principle, one could subtract the baseline background-galaxy contribution from the raw JAGB counts, resulting in slightly brighter mean and median values for the JAGB distribution, which would bring them into even closer correspondence with the TRGB predicted value.
Refining this process of second-order corrections being applied to the JAGB distance scale will be taken more fully up in a future paper.

\end{document}